\documentclass[aps,prb,showpacs]{revtex4}

\usepackage{graphicx}
\DeclareGraphicsExtensions{.eps}

\newcommand{\fFMR} {$f^{(0)}_{\rm FMR}$}

\begin{document}

\title{Magnetization precession due to a spin polarized current in a thin  
nanoelement: numerical simulation study}
\author{D.V.~Berkov, N.L.~Gorn}
\affiliation{Innovent e.V., Pr\"uessingstr. 27B, D-07745, Jena, Germany}

\date{\today}

\begin{abstract}

In this paper a detailed numerical study (in frames of the Slonczewski formalism) of 
magnetization oscillations driven by a spin-polarized current through a thin elliptical 
nanoelement is presented. We show that a sophisticated micromagnetic model, where 
a polycrystalline structure of a nanoelement is taken into account, can explain 
qualitatively all most important features of the magnetization oscillation spectra recently 
observed experimentally (S.I. Kiselev et al., {\it Nature}, {\bf 425}, 380 (2003)), 
namely: existence of several equidistant spectral bands, sharp onset and abrupt disappearance 
of magnetization oscillations with increasing current, absence of the out-of-plane regime 
predicted by a macrospin model and the relation between frequencies of so called 
small-angle and quasichaotic oscillations. However, a quantitative agreement with
experimental results (especially concerning the frequency of quasichaotic oscillations) 
could not be achieved in the region of reasonable parameter values, indicating that 
further model refinement is necessary for a complete understanding of the spin-driven 
magnetization precession even in this relatively simple experimental situation.
 
\end{abstract}


\maketitle

\section{Introduction}

Magnetic excitations and magnetization switching in nanostructures induced by a spin-polarized 
current (SPC), first predicted theoretically  \cite{SpInjPred} and soon discovered experimentally 
\cite{SpInjDiscov} are now one of the most intensively studied topics in solid state 
magnetism due to their importance both from the fundamental point of view and for numerous 
possible technical applications in, e.g., microwave generators and MRAM cells (see recent 
reviews\cite{SpinElectrRev}). In the past decade a substantial progress has been achieved 
by the elaboration of analytical theories explaining the spin-transfer driven magnetization 
dynamics and by refining experimental techniques in order to enable quantitative
experimental studies of these phenomena. 

Analytical models employing the scattering matrix formalism \cite{Waintal2000} and taking
into account spin-dependent reflection and transmission of electrons, precession of electron 
spins and the structure of Fermi surfaces for concrete materials 
\cite{Stiles2002a,Stiles2002b} have been developed which allow the calculations of 
the spin-transfer torque arising due to the current crossing the normal metal-ferromagnet 
interface. The derivation of the density operator corresponding to the normal metal spacer 
(N) in a FM/N/FM structure combined with a two-current (for spin-up and spin-down electrons) 
model enabled the evaluation of a spin torque in a complete trilayer structure 
\cite{Slonczewski2002,Manshot2004} which was successfully tested and extended using 
the Boltzmann equation formalism \cite{Xiao2004}. This way an asymmetric dependence
of the spin torque on the angle between the two FM layers could be derived 
\cite{Slonczewski2002,Stiles2002b,Manshot2004} whereby the parameters governing this
asymmetry depend on the conductivities and geometry of the multilayer device 
under study. A large effort has also been made to establish the relation between the
spin-transfer phenomena, interlayer exchange coupling and Gilbert damping in
multilayer structures (see 
Ref. \onlinecite{Tserkovnyak2002,Tserkovnyak2003a,Tserkovnyak2003b, Heinrich2003} 
and citations therein).

At the same time first high quality experimental results were obtained by studying 
magnetization oscillations induced by a spin-polarized current in various systems. 
Magnetization dynamics in nanopillars was investigated in 
Ref. \onlinecite{Kiselev2003,Kiselev2004,KiselevMMM2004,Fuchs2004,Covington2005}. 
Detailed studies of the steady-state magnetization oscillations were reported also 
for the point-contact geometry \cite{Rippard2003,Rippard2004a,Rippard2004b}. In these 
papers several different precession modes were identified and the dependencies
of the precession frequency and microwave oscillation power on the external field  
and dc current strength were investigated.

Quantitative comparison between theory and experiment remains, however, a delicate issue
due to fairly complicated remagnetization processes involved. Experimentally accessible
multilayered nanostructures have thicknesses in the range of several nanometers and 
the lateral sizes about 100 nm and larger which is far above the critical size for 
the single-domain behaviour. For this reason one expects the formation of relatively
complex domain structures during the magnetization switching, especially taking 
into account that for most equilibrium magnetization states the spin-induced torque 
has different directions for different parts of the nanoelement (see, e.g., 
the discussion in Ref. \onlinecite{Miltat2001}). Indirect evidences for such a domain-mediated
switching can be found already in first semiquantitative experimental studies 
\cite{SpInj_MultiDomSwitch}, where a quasistatic switching of a nanoelement by 
changing the external field or the electric current strength was investigated. 
Recent experiments have confirmed that a macrospin (single-domain) approximation can 
not explain several important observations concerning the magnetization dynamics in 
nanopillars and point contacts \cite{Kiselev2003,Rippard2004a,Covington2005}, 
although the macrospin model was able to predict some of its qualitative 
aspects like the existence of a steady-state precession regime by itself 
(for the most detailed study in frames of a macrospin approach see 
Ref. \onlinecite{Sun2000}).

For this reason the full finite-element micromagnetic modelling is required in order
to understand whether the discrepancies between theoretical predictions and experimental
observations are due to a poor theoretical understanding of a basic physics involved
or because the formation of complicated domain structure significantly alters the
experimentally measurable system properties. Until the present time very few such studies 
have been carried out. In the pioneering paper \cite{Miltat2001} devoted to this
subject the formation of complicated magnetization states already in a nanoelement 
with lateral sizes as small as $125 \times 125$ nm was found. In the latter detailed
study \cite{Li2003} of the magnetization dynamics of a $64 \times 64$ nm square with
the thickness 2.5 nm it was predicted that the behaviour of this element is nearly 
single-domain. However, the results of Ref. \onlinecite{Li2003} turned out to be 
at least partially incorrect for the parameters of magnetic materials typically 
used experimentally: it was shown \cite{Berkov2005a,Berkov2005b} that for a nanoelement
with this thickness significant deviations from a single-domain behaviour appear already 
for lateral sizes $b \approx 30$ nm and a completely chaotic magnetization pattern was 
found for $b \ge 60$ nm, which clearly emphasizes the importance of full-scale 
micromagnetic simulations.

Up to our knowledge, systematic micromagnetic studies concerning the spin-injection
driven magnetization dynamics have been performed only for a SPC induced noise in 
spin-valves \cite{Zhu2004IEEE}. In addition, in several short reports 
\cite{Zhu2004JAP,Lee2004} some qualitative aspects of the magnetization dynamics 
of elliptical nanoelements (concerning mainly the transition from a homogeneous to
a non-coherent magnetization precession) were discussed. 

In this paper we present a systematic study of the magnetization dynamics of 
an elliptical nanoelement under the influence of a spin-polarized current. The paper is 
organized as follows: In Sec. \ref{sec:NumSimMeth} we outline our simulation methodology,
paying special attention to the justification of our choice of the simulation parameters.
In Sec. \ref{sec:NumSimRes} we present our simulation results, starting with a 'minimal'
micromagnetic model and carrying on with adding the effects of a polycrystalline structure
and thermal fluctuations. In the same section we analyze also the influence of the 
exchange stiffness constant on the magnetization dynamics, keeping in mind that
among various magnetic parameters of a ferromagnetic material the experimental 
determination of this constant is the most difficult task. In Sec. \ref{sec:Discuss}
we compare our results first with numerical simulations of other groups and then
with the experimental studies of the microwave oscillations in such a nanoelement.
At the end of this last Section we discuss how the factors which were not included
in our model could affect the magnetization dynamics of the system under study.

\section{Numerical simulations: methodology and choice of material parameters}
\label{sec:NumSimMeth}

Numerical simulations were carried out using our package MicroMagus \cite{MicroMagus}, 
whereby the spin injection was included in form of the Slonczewski torque 
$\Gamma = (a_J/M_S) \cdot [{\bf M} \times [{\bf M} \times {\bf S}]]$ (${\bf S}$ is 
the spin polarization direction of the current through a nanoelement). The time evolution
of the magnetization configuration was obtained by integrating the Landau-Lifshitz-Gilbert 
equation of motion for the system magnetization using the modified Bulirsch-Stoer algorithm. 
This method includes an adaptive step size control, what is especially important when 
simulating dynamics of a strongly non-homogeneous magnetization configuration.

Keeping in mind the intended comparison of our simulation results with those obtained 
experimentally, we have made an effort to choose the system parameters as close as 
possible to their values for experimentally studied nanoelements. Among these experimental
studies we have selected the publication of the Cornell group \cite{Kiselev2003}, 
where ones of the qualitatively most interesting results on Co/Cu/Co system were reported. 

An elliptical element with the geometry corresponding to that used in Ref. \onlinecite{Kiselev2003},
i.e., with lateral sizes $130 \times 70$ nm and thickness $d = 3$ nm was simulated. 
The element was discretized in plane using the mesh size $2.5 \times 2.5$ nm (we have checked 
that further mesh refinement did not lead to any noticeable changes in the simulation results).

The choice of the magnetic material parameters requires special justification. 

The saturation magnetization of thin Co layers, as it is well known, may be different from 
that of the bulk material. For this reason we have used the saturation magnetization 
$4 \pi M_S = 10$ kG as measured in Ref. \onlinecite{Kiselev2003} on a Co/Cu multilayer
system where each Co layer was 3 nm thick (see below also our analysis concerning the
influence of the saturation magnetization value on the final results).

The value of the exchange stiffness constant $A$ for a bulk Co and thin Co film is also 
a subject of controversial discussions. Some earlier measurements have lead 
to the values in the range $A = (1 - 2) \times 10^{-6}$ erg/cm (see Ref. 
\onlinecite{Stoner1950} and citations in Ref. \onlinecite{Doering1966}, pp. 368-381). In 
latter experiments \cite{Shirane1968} performed using the inelastic neutron scattering 
on bulk Co crystals and recent measurements \cite{Michels2000} carried out on thin 
Co films using the elastic small-angle neutron scattering significantly higher 
values $A = (2.8 - 3.1) \times 10^{-6}$ erg/cm have been found. For this reason we 
have chosen the value $A = 3.0 \cdot 10^{-6}$ erg/cm as a basis and have also 
studied how the smaller values of $A$ would affect the results.

The treatment of the Gilbert damping (represented in the standard version of the 
LLG-equation by the double vector product $[{\bf M} \times [{\bf M} \times {\bf H^{eff}}]]$ 
multiplied by a constant factor $\lambda$) is at present a subject of an intensive research 
and is also closely related to the problem of a correct choice of the spin torque term.
First, even in the absence of an externally driven spin-polarized current the value 
of the dissipation constant in thin film systems may substantially deviate from its 
value for the corresponding bulk material (see a short survey of experimental data 
collected in Ref. \onlinecite{Tserkovnyak2002}). This effect is usually attributed 
to a spin transfer (and subsequent relaxation) from a ferromagnet into a normal metal 
layer \cite{Tserkovnyak2002} and/or to a similar process leading to a "dynamic exchange
coupling" between two FM layers \cite{Tserkovnyak2003a}. In addition, it was shown
\cite{Tserkovnyak2003b} that due to the dependence of the spin pumping and the 
spin accumulation on the {\it angle} between the magnetization of FM layers in 
a FM-N-FM systems the damping depends on the instantaneous magnetization configuration 
of FM layers.

The same mechanisms are responsible for a non-trivial dependence of the magnitude
of the spin torque ${\bf \Gamma}$ on the angle $\theta$ between the FM layer 
magnetizations (the trivial part of this dependence, $\Gamma \sim \sin\theta$, is 
contained in the double vector product $[{\bf M} \times [{\bf M} \times {\bf S}]]$). 
Various theoretical approaches \cite{Slonczewski2002,Bauer2003a,Xiao2004} have led to 
the conclusion that in a symmetric spin valve the functional dependence of 
the spin torque on $\theta$ can be cast into the form 
$\Gamma(\theta) \sim P \Lambda^2 \sin\theta/[(\Lambda^2+1)+(\Lambda^2-1)\cos\theta]$,
where $P$ is proportional to the difference between the resistances for the electrons
with opposite spin directions and $\Lambda$ may be expressed as a function of a total
resistance and conductance of a non-magnetic metallic spacer. For a non-symmetric 
spin valves more complicated formulae were derived \cite{Xiao2004}.

Both topics discussed above - (i) dependence of the dissipation parameter $\lambda$ on the 
average and local characteristics of a multilayer and (ii) a complicated dependence of 
a spin torque magnitude on the angle between the magnetization vectors in adjacent layers - 
can and should in principle be incorporated into micromagnetic simulations. However, the study 
of the influence of these factors on the magnetization dynamics is for obvious reasons
a separate and difficult problem. In our opinion, this problem should be addressed {\it after} 
the magnetization dynamics driven by a spin-polarized current is sufficiently well understood 
for the model where both the damping parameter and the coefficient before the Slonczewski 
torque term are assumed to be constant. As we shall see below, this is still not the case 
if one goes {\it beyound} the macrospin approximation, i.e., when the complicated 
magnetization structures arising in nanoelements under realistic conditions are taken into 
account. For this reason we have decided to put both $\lambda = Const$ and $a_J = Const$ 
(so that $\Gamma(\theta) \sim \sin\theta)$ in order to find out (i) which qualitative 
features of the experimentally observed steady state precession and switching of magnetic
nanoelements can be explained within this simplified treatment of the spin-transfer
induced torque and (ii) whether this model can provide quantitative agreement with
experimental data for reasonable parameter values. In future, the comparison of our results 
presented here with simulations employing more sophisticated spin torque models
will allow the unambiguous identification of dynamic phenomena arising due to 
the abovementioned non-trivial properties of the Gilbert damping and the complicated
$\theta$-dependence of the spin torque magnitude.

Within the approach outlined above we are left with two problems: how to choose the value
of the Gilbert damping $\lambda$ and the proportionality coefficient between 
the experimentally measured current values $I$ and the Slonczewski term magnitude $a_J$
for our system. We are not aware of any independent measurements of $\lambda$ for 
the experimental situation similar to that studied in Ref. \onlinecite{Kiselev2003}, 
which are reliable enough to be used as a simulation input. Hence we have adopted the following 
strategy to choose this parameter. First, it is well known from general considerations 
and confirmed by numerical simulations \cite{Zhu2004IEEE} that the current threshold 
$I_{\rm min}$ for the onset of the steady state magnetization oscillations growth 
linearly with $\lambda$. At the same time, we have found that the critical current 
value $I_{\rm max}$ for which the oscillations disappear is nearly independent on 
the damping. For this reason we have chosen the damping which enabled us to reproduce 
a ratio of the maximal and minimal currents $I_{\rm max}/I_{\rm min}$ for which 
a significant microwave power was observed experimentally \cite{Kiselev2003}. This
procedure lead to the value $\lambda \approx 0.04$. This damping is significantly higher 
than the usually reported values of $\lambda$ in Co ($\lambda \sim 0.01 - 0.02$) which
might indicate that the effects discussed above lead (on average) to 
a substantial enhancement of the damping in the system under study.

Establishing of the relation between the total current $I$ (or the current density $j$)
and the spin torque magnitude $a_J$ is required for the calculation of the current 
induced magnetic field (Oersted field). Following the discussion above, we did not try
to calculate the corresponding proportionality coefficient using various available
theoretical models. Instead, we have again used for this purpose the value of the 
current threshold $I_{\rm min}$ reported in Ref. \onlinecite{Kiselev2003}. Namely, we have
calculated the Oersted field for the critical current value $I_{\rm min}$ 
(which is $\approx$ 2 mA for $H_{\rm ext} = 2$ kOe in Ref. \onlinecite{Kiselev2003}) 
corresponding to the onset of steady state oscillations. Afterwards, keeping the Oersted 
field constant, we have increased the value of $a_J$ until the steady state
precession emerged thus obtaining the critical value of the spin torque $a_J^{\rm cr}$.
The ratio between $I_{\rm min}$ and $a_J^{\rm cr}$ was then used as a constant 
proportionality coefficient between $a_J$ and $I$ (we need the latter to calculate 
the Oersted field for the known geometry) when we increased the spin torque 
value $a_J$ in our studies.

We have chosen the coordinate system with $0xz$ plane coinciding with the element plane and
$0x$ axis parallel to the long ellipse axis. All simulation results presented 
here were obtained for the external field ${\bf H}_0$ with components $H_{0x} = 2000$ Oe, 
$H_{0y} = 20$ Oe, $H_{0z} = 100$ Oe. The $H_{0x}$-value corresponds to that used for 
several measurement series reported in Ref. \onlinecite{Kiselev2003}, the small in-plane 
deviation $H_{0z}$ was introduced to mimic a corresponding small angle between the long 
ellipse axis and the external field \cite{Kiselev2003}, and the very small 
out-of-plane component $H_{0y}$ was introduced to remove an artificial numerical 
degeneracy present when the field is applied exactly in the layer plane. The spin 
polarization vector ${\bf S}$ was chosen to be antiparallel to the external field 
thus assuming that the magnetization direction of the polarizing (lower) layer 
coincides with the external field (the upper layer magnetization dynamics is driven by
the electrons reflected from the polarizing layer so that their average spin 
polarization is opposite to that of the lower layer).

For a direct comparison of simulation results with experimental data we need 
to calculate the spectral power of the resistance oscillations. To do this, we make an 
assumption that in the CPP-geometry (current perpendicular to plane) the resistance 
variation due to the magnetization dynamics for the current flowing through 
the $i$-th discretization cells of the free (magnetization direction 
${\bf m}_i^{\rm free}$) and fixed (${\bf m}_i^{\rm fix}$) layers can be written
as $\Delta R_i = \Delta R_{\rm max}(1 - \rm cos\theta_i)/2 = 
\Delta R_{\rm max}(1 - ({\bf m}_i^{\rm free} {\bf m}_i^{\rm fix}))/2$. We note in passing
that at present more complicate dependencies of the magnetoresistance on the angle
between the adjacent layer magnetizations are discussed (in addition to the papers
mentioned above, see, e.g., Ref. \onlinecite{Manshot2004}). However, at this stage of 
our research where we have to use the angular-independent spin torque magnitude $a_J$, 
we prefer to use the standard $\cos$-like dependence written above thus treating 
both phenomena (magnetoresistance and spin torque) in a self-consistent manner.

For the device under consideration the variation of the resistance due to the magnetization
precession is much lower than its value in the absence of magnetoresistive effects. Using 
this simplifying assumption, the total time-dependent resistance variation of our discretized
micromagnetic model treated as a system of $N_c$ parallel resistors can be expressed as 
a function of the fixed layer magnetization direction ${\bf m}^{\rm fix}$ (this layer is 
assumed to be homogeneously magnetized) and the {\it average} magnetization direction 
$\langle {\bf m}^{\rm free} \rangle$ of the free layer as

\begin{equation}
\label{MagResEq1}
\Delta R_{\rm tot}(t) = {\Delta R_{\rm max} \over 2 N_c} 
\{1 - ( {\bf m}^{\rm fix} \cdot \langle {\bf m}^{\rm free} \rangle ) \}
\end{equation}

Taking into account that the in-plane magnetization direction of the fixed layer for 
the geometry under study coincides with in-plane external field direction (given by 
the unit vector ${\bf e}^h = {\bf H}_0/H_0$) and that the out-of-plane component 
of ${\bf m}^{\rm fix}$ is vanishingly small, we finally obtain the formula

\begin{equation}
\label{MagResEq2}
\Delta R_{\rm tot}(t) = {\Delta R_{\rm max} \over 2 N_c} 
\{1 - (e^h_x \langle m_x^{\rm free} \rangle + e^h_z \langle m_z^{\rm free} \rangle )\}
\end{equation}
which enables a direct calculation of the magnetoresistance oscillation spectrum
from the known time dependencies of the average in-plane magnetization projections of
the free layer.

For further references we also note that the frequency of the homogeneous FMR-mode of the
extended thin film with magnetic parameters and under external conditions given above 
would be \fFMR $= (\gamma/2\pi) \cdot (H_0(H_0 + 4\pi M_S))^{1/2} \approx 13.7$ GHz. 

\section{Numerical simulations: Results and discussion}
\label{sec:NumSimRes}

\subsection{Magnetization dynamics for $T = 0$: a 'minimal' micromagnetic model
with the reference parameter set}
\label{subsec:MinModT0}

We begin with the description of the steady state precession without thermal noise. 
In this case there exist for a fixed external field and increasing current (increasing $a_J$ 
in our formalism) a sharp transition from a stable magnetization state (time-independent
magnetization configuration) to a steady-state regime with regular oscillations. 
In this subsection we consider a 'minimal' micromagnetic model, where the polycrystalline 
structure of a nanoelement is neglected, so that no random magnetic anisotropy of crystal 
grains is taken into account.

\begin{figure}[tbhp] 
\centering
{\includegraphics
[scale=0.8, bb=5cm 1cm 15cm 28cm]
{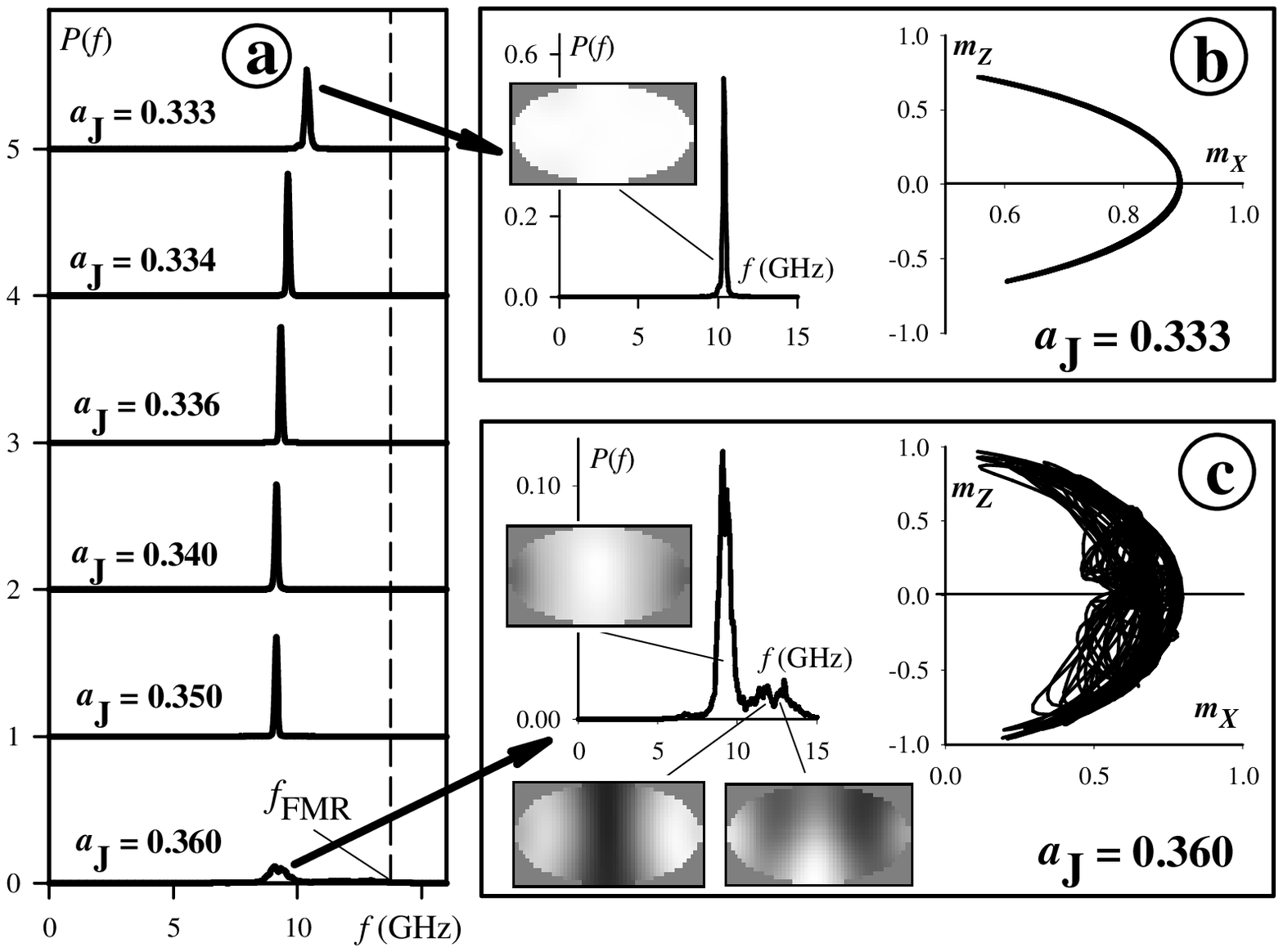}}
\caption
{
Magnetization dynamics for an elliptical element with a 'reference' parameter set near the 
oscillation onset threshold. Panel (a): power spectra of the $m_z^{\rm av}$-component for
various $a_J$ values as indicated in the figure. Panel (b): magnetization dynamics for 
$a_J = 0.333$ represented by the enlarged $m_z^{\rm av}$-oscillation spectrum and 
the projection of the average magnetization trajectory on the coordinate plane $0xz$ 
(coinciding with the nanoelement plane). Panel (c): the same as on (b) for $a_J = 0.360$ - 
immediately after the transition to a quasichaotic oscillation regime. Grey-scale maps 
represent spatial distributions of the $m_z^{\rm av}$ oscillation power for corresponding 
spectral peaks.
}
\label{fig:RegOscillMinMod} 
\end{figure}

We start the analysis of our simulation results from the regular oscillation regime, which 
is characterised by a nearly homogeneous magnetization configuration during the precession.
For the nanoelement parameters ($4 \pi M_S = 10$ kG, $A = 3.0 \cdot 10^{-6}$ erg/cm and
$\lambda = 0.04$) and the external field (${\bf H_0} = (2000,20,100)$ Oe) given above, the 
current threshold for the oscillation onset in our "minimal" model was found to be 
$a_J^{\rm cr} = 0.3310(5)$. When $a_J$ is increased above this value, a very fast growth 
of the oscillation amplitude was observed. In particular, for $a_J = 0.333$ the oscillation 
amplitude of the average $z-$projection of the element magnetization $m_z^{\rm av}$ 
(the in-plane projection along the short ellipse axis, nearly perpendicular to the external 
field) was already $\Delta_z \approx 0.7$ (see Fig. \ref{fig:RegOscillMinMod}b). The regular 
oscillation regime exists in a extremely narrow $a_J$-region $0.331 \le a_J \le 0.355$. 
In this regime the spectrum of $m_z^{\rm av}$-oscillations contains only a single very narrow 
peak which frequency is $f_1 \approx 10.37$ GHz for $a_J = 0.333$ and decreases down to 
$f_1 \approx 9.14$ GHz for $a_J = 0.350$. We point out that already for $a_J = 0.333$ 
the steady-state oscillation frequency is well below the frequency of the homogeneous 
FMR-mode \fFMR, although the distribution of the oscillation power is almost 
perfectly homogeneous (Fig. \ref{fig:RegOscillMinMod}b, left panel). The reason is
that even for this $a_J$-value - which is very close to the oscillation onset threshold 
$a_J^{\rm cr}$ - the oscillation amplitude is so large that the formula 
\fFMR $= (\gamma/2\pi) \cdot (H_0(H_0 + 4\pi M_S))^{1/2}$ used to calculate \fFMR 
 (where {\it small} deviations of magnetization from its equilibrium orientation are assumed) 
is not applicable.

As mentioned above, the regular oscillation regime with a well defined limiting cycles for 
the average system magnetization (Fig. \ref{fig:RegOscillMinMod}b, right panel) 
persists up to $a_J^{\rm ch} \approx 0.355$. Further small increment of the current strength 
leads to an abrupt transition to a quasichaotic magnetization motion (whether it is truly 
chaotic in the sense of non-linear dynamics should be investigated separately; see also our 
brief discussion in Ref. \onlinecite{Berkov2005a}). An example of the magnetization 
trajectories in this quasichaotic regime is given in Fig. \ref{fig:RegOscillMinMod}c.

For currents not much larger than the transition value $a_J^{\rm ch}$ the oscillation spectra 
of the average magnetization components still exhibit a well defined structure with several 
peaks corresponding to several oscillation eigenmodes with different spatial distribution 
of the oscillation power. An example of the such a spectrum with spatial maps attributed
to different spectral peaks is given in Fig. \ref{fig:RegOscillMinMod}c (left panel). In-plane 
spatial distributions of the oscillation power were calculated from the time-dependent
trajectories of the discretization cell moments as described in Ref. \onlinecite{Berkov2005b}.
The detailed analysis of these spatial structures is beyound the scope of this paper.
Here we would like only to mention that the asymmetry of the spatial distributions for
the second and third modes is due to the presence of the Oersted field. Although this
field is much smaller than the external field, its influence for this particular case
is visible due to a relatively large amplitude of the magnetization oscillation.

Further increment of the current strength ($a_J$ value) results in the gradual decline
of the oscillation power for both in-plane magnetization components, accompanied by
the decrease of the oscillation frequency as a function of $a_J$. This behaviour is 
demonstrated in Fig. \ref{fig:ChaoticOscillMinMod}, where we present the maps of 
$m_x^{\rm av}$- and $m_z^{\rm av}$-oscillation power as functions of the frequency 
(vertical axis) and $a_J$ (horizontal axis) together with the magnetoresistance oscillation 
power computed according to Eq. (\ref{MagResEq2}). Characteristic for this dynamical 
regime is a multi-domain magnetization configuration whereby the average domain size 
decreases with increasing $a_J$. Because the average magnetization is still directed along 
the external field, a formation of such a chaotic domain structure with smaller and 
smaller domains leads on average to the increase of the stray field in the opposite 
direction, which might be one of the reasons why the oscillation frequency decreases 
with $a_J$.

It is important to note, that in contrast to $m_z^{\rm av}$, the oscillation power of the 
$m_x^{\rm av}$-component (parallel to the long ellipse axis) is concentrated in {\it two} 
frequency regions. First, there is a contribution from fast $m_x^{\rm av}$ oscillations 
at the frequency approximately twice as high as for $m_z^{\rm av}$ oscillations. The reason for
this relation between $m_x$ and $m_z$ oscillation frequencies is that for the motion type under 
study - the magnetization is precessing around (approximately) the $0x$-direction - the 
$m_x$-component moves back and forth twice during one complete oscillation cycle (see, e.g., 
right graphs in Fig. \ref{fig:RegOscillMinMod}b and c). Second, there exist a 
substantial low-frequency contribution to the oscillation power (which growth with 
increasing $a_j$) due to slow variations of $m_x^{\rm av}(t)$ which are typical for 
a quasichaotic motion. Analogous behaviour of the longitudinal magnetization component 
in similar situations was observed in 
Ref. \onlinecite{Zhu2004JAP,Zhu2004IEEE,Lee2004,Berkov2005a} (see also the discussion 
below).

Increasing $a_J$ above the value $a_J \approx 1.4$ leads to the disappearance of the chaotic
regime and to the establishing of a so-called 'out-of-plane' precession mode. The magnetization
trajectory in this mode is an ellipse with the plane slightly tilted with respect
to the nanoelement plane. The oscillation frequency abruptly increases and is now for obvious
reasons the same for $m_x^{\rm av}$ and $m_z^{\rm av}$-components (see regions marked by 
circles in Fig. \ref{fig:ChaoticOscillMinMod}). The magnetization is precessing nearly coherent
in this mode, what can be seen from a large absolute value of the out-of-plane
magnetization projection: ($m_y^{av} \approx 0.8 - 0.9$). The absolute value of the oscillation 
power in this regime is at least an order of magnitude smaller than at the beginning of the 
quasichaotic motion due a relatively small precession amplitude.

Comparison of these results with experimental data and numerical simulations of other groups
will be made in Section \ref{sec:Discuss}.

\subsection{Magnetization dynamics for $T = 0$: influence of the exchange stiffness value}
\label{subsec:ExchA2T0}

Taking into account a relatively broad region of the exchange stiffness constants reported
for bulk Co and thin Co films (see discussion at the beginning of this Section), we have 
studied how the decrease of $A$ would affect the magnetization dynamics. From 
general considerations, we expect the magnetization configuration to become 'softer' for
smaller exchange stiffness, so that the influence of a strongly inhomogeneous 
self-demagnetizing field of a flat elliptical nanoelement should lead to an 
equilibrium magnetization configuration which is less collinear compared to a system
with larger $A$. This, in turn, should result in a broader distribution of the resonance
fields for different regions of a nanoelement (different discretization cells in our 
micromagnetic model), thus leading to a slower increase of the  magnetization oscillation 
amplitude with increasing $a_J$ (above the oscillation threshold 
$a_J^{\rm cr}$). In addition, for smaller exchange the transition from a regular to
a quasichaotic oscillation regime is expected to occur earlier (for smaller $a_J$).

All these features were indeed observed in our simulations, whereby it has turned out
that from the quantitative point of view the nanoelement with the geometry under study
is quite sensitive already to moderate decrease of $A$. In Fig. \ref{fig:RegOscillA2}
we display an example of the transition from a regular to a quasichaotic behaviour for
the nanoelement with parameters identical to those given in the preceding subsection, but 
with $A = 2.0 \cdot 10^{-6}$ erg/cm. The oscillation threshold $a_J^{\rm cr} = 0.3290(5)$ 
is almost the same as for the system with the reference parameter set. However,
the spectral power of magnetization oscillations in a regular regime is substantially
smaller than for a 'reference' system. In particular, when $a_J$ exceeds the oscillation 
threshold value by $\Delta a_J = 0.0015$ ($a_J = 0.331$, see corresponding spectrum
in Fig. \ref{fig:RegOscillA2}a and trajectories in Fig. \ref{fig:RegOscillA2}b), the oscillation
amplitude is significantly smaller than for the same situation for a 'reference' system (Fig. 
\ref{fig:RegOscillMinMod}b). For this reason the oscillation frequency here 
($f \approx 12.45$ GHz) is much closer to \fFMR than for the peak shown in 
Fig. \ref{fig:RegOscillMinMod}b.

The transition to a quasichaotic motion occurs, due to the reason explained above, for 
a somewhat smaller current value $a_J^{\rm ch} = 0.338(1)$. Interestingly, the non-trivial 
mode structure appears for this system already before the transition (see spectra for 
$a_J = 0.334$ and $a_J = 0.336$ in Fig. \ref{fig:RegOscillA2}a). In addition, for the system with 
smaller $A$ only the first and third mode from of the eigenmode set identified for 
$A = 3.0 \cdot 10^{-6}$ erg/cm are visible in the power spectrum of magnetization 
oscillations just above the transition to chaos. The analysis of trajectories for 
{\it individual} discretization cells reveals that the second mode (with the intermediate 
frequency) is still present. However, the spatial structure of this mode for this smaller 
exchange value is nearly symmetric (compare oscillation power maps in 
Fig. \ref{fig:RegOscillMinMod}c and \ref{fig:RegOscillA2}c), and the mode does not manifest 
itself in the spectrum of the {\it average} magnetization dynamics, because 
the magnetization in the left and right mode localization regions oscillates in 
opposite phases.

Analysing the influence of exchange stiffness value on the overall spectral picture in 
the chaotic regime, we can see that the decrease of $A$ increases the upper boundary of
the current region where the quasichaotic regime exists (compare 
Fig. \ref{fig:CompChaoticOscill_A3_A2_fccCo}a and \ref{fig:CompChaoticOscill_A3_A2_fccCo}b). 
In addition, it can be seen that the decrease of $A$ down to $A = 2.0 \cdot 10^{-6}$ erg/cm
leads to a nearly complete disappearance of the 'out-of-plane' precession region. This
means that for large spin torque values this exchange constant is not high enough to ensure 
the collinear magnetization structure necessary for the very existence of the 
'out-of-plane' precession mode.
 
\subsection{Magnetization dynamics for $T = 0$: influence of the random
crystal grain anisotropy}
\label{subsec:RandAnisT0}

It is well known that thin sputtered Co films have a polycrystalline structure
which in the simplest case can be treated as absolutely random and characterised
by a single parameter - the average grain size $\langle D \rangle$. In all the 
simulations which results are shown below we have used $\langle D \rangle = 10$ nm.
To study a nanoelement with such a random polycrystalline structure we have applied
a standard procedure included into the MicroMagus \cite{MicroMagus} package. 
The procedure starts with the random
placement of 'growth centers' within the simulated nanoelement, whereby the number 
of growth centers is determined from the nanoelement size and the average grain size 
$\langle D \rangle$. Afterwards the crystal grains are 'build' simulating
a simple isotropic 2D growth which starts simultaneously at all growth centers and
is terminated at the locations where the adjacent grains touch each other. The routine 
stops as soon as the whole element area is covered by the grains. Finally a random 
direction of the anisotropy axes is assigned to each grain, which means that (i) for all 
discretization cells within a given grain the anisotropy axes directions are the same,
but (ii) the corresponding directions for two different grains are not correlated.

Thin Co films may have a polycrystalline structure with grains having either 
{\it fcc} or {\it hcp} crystal lattice type. Co films with thicknesses about 
several nanometers epitaxially grown on Cu substrates possess the {\it fcc} 
crystallographic structure (see, e.g., review \cite{Heinz1999} and references therein),
so it is likely that Co films sputtered on the same substrate (as those used 
in Ref. \onlinecite{Kiselev2003}) also have predominantly the {\it fcc} structure 
(see also results and discussion in Ref. \onlinecite{Langer2001}). Hence we begin 
this subsection with the analysis of simulation results for the corresponding case.

\subsubsection{Magnetization dynamics for a nanoelement with {\it fcc} crystal grains}
\label{subsubsec:RandAnisFCC_T0}

At room temperatures the {\it fcc} crystallographic modification of Co is unstable for bulk
crystals, so that measurements of the anisotropy constant can be performed only
on thin films which {\it fcc} structure is stabilized by, e.g., a suitable substrate choice
(like Cu with the 111-orientation of the substrate surface). This circumstance leads
to considerable difficulties by the anisotropy constant determination, which could be
performed mainly on epitaxially grown films (see, e.g., Ref. \onlinecite{CoFCCMagAnis}). 
The values of the first cubic anisotropy constant obtained for thin Co films grown 
at different conditions on different substrates are within the range 
$K_1^{\rm cub} = (5 - 8) \cdot 10^5$ erg/cm$^3$, so we have used the value 
$K_1^{\rm cub} = 6.0 \cdot 10^5$ erg/cm$^3$ in our simulations. 
Varying $K_1^{\rm cub}$ within the region cited above did not lead to any 
significant changes of the results.

From this anisotropy value and the standard saturation magnetization of Co 
$M_S \approx 1400$ G (which is almost the same for {\it fcc} and {\it hcp} phases) 
one can determine the reduced anisotropy constant $\beta = 2K_1/M_S^2$, also sometimes 
referred to as a 'quality factor' $Q$. The corresponding value $\beta \approx 0.6$ is 
not small (and is even larger if we substitute into the expression for $\beta$ the
reduced $M_S$ value reported in Ref. \onlinecite{Kiselev2003}), which means that 
this random anisotropy may cause substantial inhomogeneities 
of magnetization structure in a polycrystalline sample. In nanocrystalline thin films, 
however, where the grain size is of the same order of magnitude as the exchange length, 
the influence of the random anisotropy is strongly diminished, because its effect is 
'averaged out' due to uncorrelated anisotropy directions in adjacent crystallites 
\cite{Herzer1997}.

Nevertheless in the nanoelement under study (with the average grain size 
$\langle D \rangle = 10$ nm) the random cubic anisotropy leads to a noticeable
deviations of the magnetization structure from the configuration obtained for an ideal
nanoelement (without the polycrystalline structure) and thus results in an observable spread
of resonance fields for different crystallites. Qualitatively this leads to the same effect
as the decrease of the exchange constant: after passing the threshold value 
$a_J^{\rm cr} = 0.337(2)$ the amplitude of the SPC-driven precession of the average 
magnetization increases much slower, than in the absence of the random anisotropy. In
particular, for $a_J = 0.340$ we can still observe a really small-angle homogeneous 
precession with the frequency $f_{\rm hom} \approx 13.6$ GHz which is virtually equal
to \fFMR ($\approx 13.7$ GHz) - see Fig. \ref{fig:RegOscill_fccAnis}b. 

In contrast to the decrease of the exchange constant, which leads to {\it systematic} 
deviations of the magnetization configuration from a homogeneously magnetized state under 
the influence of the self-demagnetizing field, a crystal grain anisotropy leads to 
{\it random} deviations of the magnetization. For this reason the mode structure after 
the transition to the quasichaotic precession (Fig. \ref{fig:RegOscill_fccAnis}c) looks 
entirely different when compared with the case of the reduced exchange constant
(Fig. \ref{fig:RegOscillA2}c). And finally, the analysis of the oscillation spectra in 
the quasichaotic region of currents for the nanoelement with a random polycrystalline
structure (Fig. \ref{fig:CompChaoticOscill_A3_A2_fccCo}c) shows that the presence of such 
an anisotropy slightly expands the existence region of quasichaotic oscillations
(compare with Fig. \ref{fig:CompChaoticOscill_A3_A2_fccCo}a) and significantly decreases 
the spectral power of 'out-of-plane' oscillations. The latter effect is due to the same 
reason as for the case of a reduced exchange - random anisotropy disturbs the homogeneous
magnetization configuration required for the existence of an 'out-of-plane' precession.

\subsubsection{Magnetization dynamics for a nanoelement with {\it hcp} crystal grains}
\label{subsubsec:RandAnisHCP_T0}

It has been shown experimentally \cite{Langer2001} that although crystallites 
of thin Co films sputtered on Cu/111 substrates have mostly the {\it fcc} structure, there
exist a non-negligible fraction of grains possessing a 'normal' {\it hcp} structure of
bulk Co crystals. For this reason we have also studied the magnetization dynamics
of a polycrystalline nanoelement with {\it hcp}-grains having the same average size 
$\langle D \rangle = 10$ nm.

Magnetic anisotropy of the {\it hcp} modification of Co can be treated for our purposes as
a uniaxial one with only the first anisotropy constant 
$K_1^{\rm cub} = 4.5 \cdot 10^6$ erg/cm$^3$ 
(see, e.g., Ref. \onlinecite{ChikazumiBook}) taken into account. 
The reduced anisotropy constant $\beta \approx 4.6$ calculated from this $K_1$ value and 
the same saturation magnetization $M_S \approx 1400$ G is so large that the random 
anisotropy plays a crucial role even in 
a sample with such small crystallites. This can be seen very well in 
Fig. \ref{fig:CompDiffRealiz_hcpCo}, where we show maps of the oscillation power for
polycrystalline nanoelements with {\it hcp} grains. 
The four maps (a)-(d) were computed for 'samples' with {\it identical} macroscopic 
parameters ('reference' parameter set and one and the same value of the random uniaxial 
anisotropy of crystallites $K_1^{\rm cub} = 4.5 \cdot 10^6$ erg/cm$^3$), but using 
{\it different} realizations of a random polycrystalline structure. The most striking 
feature of the spectra shown in Fig. \ref{fig:CompDiffRealiz_hcpCo} is that they are
qualitatively different for different 'samples'. In particular, we can recognize 
both (i) a large spread of the regular oscillation frequencies (narrow peaks for  
small $a_J$ lie in a wide range $\approx 10 - 20$ GHz) and (ii) very different width and 
$a_J$-dependencies of broad spectral bands in the region of a quasichaotic behaviour.
In addition, we observe large frequency jumps by the transition from a regular to 
a quasichaotic regime (see, e.g., realizations (a) and (b)). This is in strong contrast
to the behaviour of a nanoelement with an absent or moderate anisotropy, where the regular
oscillation frequency before the transition to a quasichaotic regime was 'inherited' 
by a main spectral peak of quasichaotic oscillations just after the 
transition.

Further insight into the magnetization dynamics of a polycrystalline nanoelement with
{\it hcp} grains can be gained by the analysis of the magnetization trajectories and spatial
distribution of the oscillation power. 3D trajectories of the average system magnetization
for the 'sample' marked as (b) in Fig. \ref{fig:CompDiffRealiz_hcpCo} at the $a_J$ values
before ($a_J = 0.50$) and immediately after ($a_J = 0.54$) the transition from a regular
to a quasichaotic regime are shown in Fig. \ref{fig:3Dtraj_SpecMxMz_hcpCo} together with 
$m_x^{\rm av}$ and $m_z^{\rm av}$ power spectra and spatial maps of $m_x^{\rm av}$ and 
$m_z^{\rm av}$ oscillation power. 

First of all we note that due to a strong influence of the {\it hcp} anisotropy with randomly
oriented anisotropy axes in various grains, the average nanoelement magnetization does
not precess anymore approximately around the $0x$ axis (long ellipse axis) even in the 
regular regime. This is clearly demonstrated in Fig. \ref{fig:3Dtraj_SpecMxMz_hcpCo}a:
the limiting cycle in the form of a narrow bent torus is strongly tilted with respect
to the $0x$-axis. Such an asymmetric trajectory lead to the appearance of a strong 
spectral peak at the basic precession frequency also in the oscillation spectrum of
the longitudinal magnetization component $m_x^{\rm av}$ (compare with spectral maps
for an element without a random anisotropy in Fig. \ref{fig:ChaoticOscillMinMod}a,b).

The next important feature is a strongly irregular spatial distribution of the oscillation
power which is especially pronounced for the $m_x^{\rm av}$-component (grey-scale map 
in Fig. \ref{fig:3Dtraj_SpecMxMz_hcpCo}b). It is evident that this power distribution is 
determined by the random polycrystalline structure; this circumstance is again in a strong 
contrast to the case of an absent or moderate {\it fcc} anisotropy. By the transition to 
a quasichaotic regime this spatial distribution changes drastically (compare corresponding 
maps in Fig. \ref{fig:3Dtraj_SpecMxMz_hcpCo}b and \ref{fig:3Dtraj_SpecMxMz_hcpCo}e) thus 
causing a relatively large frequency jump by this transition. For example, the difference 
between the positions of the peaks corresponding to basic oscillation frequencies in 
Fig. \ref{fig:3Dtraj_SpecMxMz_hcpCo}b and \ref{fig:3Dtraj_SpecMxMz_hcpCo}e 
is $\approx 3.5$ GHz.

Further discussion of these results is again postponed till the next Section.

\subsection{Magnetization dynamics for finite temperature: influence of thermal
fluctuations}
\label{subsec:SimResT300}

Simulations of the magnetization dynamics taking into account thermal fluctuations
were performed by adding the 'fluctuation field' ${\bf H}^{\rm fl}$ to the total 
effective field in the LLG equation of motion. We have used components of 
${\bf H}^{\rm fl}$ with zero means and $\delta$-correlated in space and time

\begin{equation}
\label{Hfl_Corr}
\langle H_{\psi, i}^{\rm fl}(0) H_{\xi, j}^{\rm fl}(t) \rangle \ = 
2D \delta_{\psi,\xi} \delta_{i,j} \delta(t)
\end{equation}
where the noise power $D = (\lambda/1 + \lambda^2) \cdot (kT/\gamma \mu)$ depends on 
the system temperature $T$, damping $\lambda$ and the total magnetic moment of 
a discretization cell $\mu = M_S \Delta V$. In principle it is known that 
the non-trivial correlation properties of the random noise appear in numerical 
micromagnetics due to the discretization of a formally continuous problem.
We refer the interested reader to Ref. \onlinecite{Berkov2005b,Berkov2004} for 
the detailed discussion about these properties and their possible influence on the 
magnetization dynamics. Here we only note that we have checked the issues
mentioned in Ref. \onlinecite{Berkov2005b,Berkov2004} to ensure that the approximation 
(\ref{Hfl_Corr}) is sufficiently good for our situation. 

Another problem arising by numerical simulations at finite temperatures is whether
one should change the macroscopic magnetic parameters of the material under study when
doing such simulations (see Ref. \onlinecite{Grinstein2003} and citations therein for 
the recent discussion). Taking into account that we use fairly small discretization 
cells and simulate the system behaviour at room temperature ($T = 300$ K) which is well 
below the Curie temperature of common magnetic materials, we have left the macroscopic
parameters unchanged. However, we note that the arguments presented in 
Ref. \onlinecite{Grinstein2003} make further studies in this direction worthwhile.

We also point out that the inclusion of the fluctuation field ${\bf H}^{\rm fl}$ with
correlation properties given by (\ref{Hfl_Corr}) converts the ordinary differential 
LLG-equation into a stochastic one. Due to random functions appearing on the right-hand
side of such equations numerical methods for their solution have, in general, a much 
lower accuracy order than their counterparts for ordinary differential equations 
\cite{KloedenBook}. For this reason a much smaller time step is required to integrate
the LLG-equation with the prescribed accuracy at $T > 0$. In our particular case, for 
$T = 300$ K and discretization cells with a relatively small size (note that the noise 
power in Eq. (\ref{Hfl_Corr}) is inversely proportional to the cell volume), the time step 
had to be decreased by more than order of magnitude. Hence for $T = 300$ K we have 
performed simulations only during the physical time of $\approx 50$ ns (after establishing 
the steady state oscillation regime). This circumstance explains why the statistical errors 
in spectra shown in Fig. \ref{fig:AllRegimes_T300} are considerably larger than for those 
analysed in previous subsections (for $T = 0$), especially for a quasichaotic oscillation 
regime. 

Example results of our simulations including thermal fluctuations are displayed in 
Fig. \ref{fig:AllRegimes_T300}, where we present the analysis of the magnetization dynamics
of a polycrystalline nanoelement with a 'reference' parameter set and {\it fcc} Co grains.

Comparing these data with the results shown in Fig. \ref{fig:RegOscill_fccAnis} and 
\ref{fig:CompChaoticOscill_A3_A2_fccCo}c we can see, first of all, that thermal fluctuations
have a qualitative influence on the regular precession regime (the region of regular 
oscillations is marked also in Fig. \ref{fig:AllRegimes_T300}). On the one hand, the
$a_J$-region where corresponding oscillations exist, is much broader than for $T = 0$:
already for $a_J = 0.2$ the $m_z^{\rm av}$ oscillation power is twice as large as in the
absence of a spin current. On the other hand, the width of corresponding spectral lines
drastically increases. For $T = 0$ the linewidth of $m_z^{\rm av}$-oscillation spectra
for a regular precession was always $\Delta f < 20$ MHz, whereby this number represents 
an upper limit posed on $\Delta f$ by the finite duration of our simulations. Inclusion
of a thermal noise leads to strong $a_J$-dependent broadening of the spectral lines:
at room temperature in the regular regime $\Delta f$ {\it decreases} with $a_J$ starting 
from the value $\Delta f \approx 1.5 (\pm 0.2)$ GHz for the spectrum of spontaneous 
oscillations (without a spin current, $a_J = 0$) down to $\Delta f \approx 0.4 (\pm 0.1)$ 
GHz for $a_J = 0.3$.

The second effect of the thermal noise is the strong decrease of the 'out-of-plane' 
oscillation power. Comparison of Fig. \ref{fig:AllRegimes_T300}b and 
\ref{fig:CompChaoticOscill_A3_A2_fccCo}c shows that in the presence of thermal fluctuations
this power is smaller by more than one order of magnitude. The reason is the same as 
explained above by the comparison of results for a 'minimal model' and its
modifications presented in Fig. \ref{fig:CompChaoticOscill_A3_A2_fccCo}.

And finally we note that according to our simulations, thermal noise does not change 
the frequencies of spectral peaks in a region where a quasichaotic precession regime 
is well established (starting from $a_J \approx 0.38$ for the system studied here). 
Comparison of corresponding oscillation spectra reveals that the frequencies of 
both peaks remains the same in frames of statistical errors.

\section{Discussion}
\label{sec:Discuss}

\subsection{Comparison with numerical simulation results of other groups}
\label{subsec:CompOtherNumSim}

Full-scale micromagnetic simulations of a magnetization dynamics driven by a spin-polarized 
current have been performed, up to our knowledge, only in a few papers, which are partially
mentioned in the Introduction. 

The first systematic study of the SPC-driven magnetization dynamics was performed in 
Ref. \onlinecite{Zhu2004IEEE} where a noise induced in a double-layer system with the 
geometry corresponding to a CPP spin valve head was investigated. Several features of 
the spin-current induced precession observed in Ref. \onlinecite{Zhu2004IEEE} (fast increase 
of the magnetization oscillation power after the appearance of a steady state precession, 
linear dependence of the current threshold value on the damping constant $\lambda$,
uniform magnetization precession for low current values and chaotic magnetization motion
for large currents, significant low-frequency contribution to the oscillation power for 
large currents) are qualitatively similar to our results. However, a quantitative comparison 
is not possible, because geometry and material parameters used in Ref. \onlinecite{Zhu2004IEEE} 
are very different from ours.

The steady-state magnetization precession in an elliptical single-layer nanoelement 'made of' 
Permalloy (${\rm Ni}_{81} {\rm Fe}_{19}$) was studied for the first time in the short report 
\cite{Zhu2004JAP}. A polycrystalline structure of a nanoelement was neglected taking into
account an extremely small anisotropy of Py grains. Both regular and quasichaotic precession 
regimes were found in Ref. \onlinecite{Zhu2004JAP} for an element with lateral sizes 
$100 \times 50$ nm. The increase of the oscillation amplitude beyound the threshold of 
the steady-state precession onset observed in Ref. \onlinecite{Zhu2004JAP} is somewhat 
slower than in our simulations. The current region (normalized 
on the precession onset value) where a regular precession regime exists in 
Ref. \onlinecite{Zhu2004JAP} is broader than for our nanoelements 
without a polycrystalline structure. Both quantitative discrepancies can be explained by
the smaller exchange constant $A = 1 \times 10^{-6}$ of Permalloy (compared to our system)
and a smaller element size used in Ref. \onlinecite{Zhu2004JAP}.

Simulations most closely related to our studies were done in Ref. \onlinecite{Lee2004}, where
the magnetization dynamics of a single-layer elliptical element with the same geometry
as used by us was performed. Lee et al. \cite{Lee2004} did not study how the magnetisation 
dynamics depends on the nanoelement magnetic parameters and did not carry out 
a systematic analysis of the oscillation spectra, but it is nevertheless possible 
to compare some results from Ref. \onlinecite{Lee2004} with our findings. 

First of all we note that Lee et al. simulated a nanoelement with the standard saturation 
magnetization value of the bulk Co $M_S = 1400$ G. Due to this $M_S$-value they could
obtain a satisfactory agreement between the oscillation frequency in a regular precession
regime (called a "small-amplitude signal" in Ref. \onlinecite{Lee2004} and 
\onlinecite{Kiselev2003}) obtained in their simulations and experimentally \cite{Kiselev2003}. 
For example, for the external field $H_0 = 2$ kOe simulations performed in 
Ref. \onlinecite{Lee2004} produce $f_{\rm sim} \approx 16.5$ GHz (see Fig. 3b in 
Ref. \onlinecite{Lee2004}), whereby experiment of the Cornell group gives 
\cite{Kiselev2003} $f_{\rm exp} \approx 16$ GHz. This agreement is a natural consequence
of a higher $M_S$ value used in in Ref. \onlinecite{Lee2004} compared to our simulations and
the $M_S$-value measured for a sample used in Ref. \onlinecite{Kiselev2003}: 
the Kittel FMR-frequency obtained for a thin film with $M_S = 1400$ G at $H_0 = 2000$ Oe is 
$f^{(0)}_{\rm FMR} \approx 17.5$ GHz. An slightly smaller value simulated 
in Ref. \onlinecite{Lee2004} is most probably due a slightly non-linear precession character
(see our discussion of results presented in our Fig. \ref{fig:RegOscillMinMod}).

We have also performed simulations with $M_S = 1400$ G 
(results are not shown here) and have obtained for a nanoelement with the same geometry
a regular precession with the frequency $f_{\rm sim} \approx 17.2$ GHz, which is very 
close both to the value obtained in Ref. \onlinecite{Lee2004} and that calculated from the Kittel 
formula. However, the quasichaotic regime for an element with such a high saturation 
magnetization starts by our simulations at the frequency $\approx 16$ GHz what is 
approximately 3 times larger than the experimentally observed value (see Fig. 1c and 
1f in Ref. \onlinecite{Kiselev2003}). In this connection we note, that although
the frequencies for a quasichaotic regime are not given in Ref. \onlinecite{Lee2004}, 
the transition scenario from a regular to a quasichaotic precession shown in Fig. 3c 
from Ref. \onlinecite{Lee2004} is similar to ours (see, e.g., 
Fig. \ref{fig:RegOscillMinMod} in this paper), so that
we expect that the frequencies at the beginning of a quasichaotic motion for simulations 
performed in Ref. \onlinecite{Lee2004} are also approximately as high as those 
for regular oscillations.

In addition, the relation of the maximal to the minimal current of the range where the
quasichaotic regime exists for a nanoelement with $M_S = 1400$ G is also much larger 
than the value $I_{\rm max}/I_{\rm min} \approx 3$ deduced from the data presented in 
Fig. 1f of Ref. \onlinecite{Kiselev2003}. The relation $I_{\rm max}/I_{\rm min}$ increases, 
because the orientation (on average) of the magnetization along the external field is 
preserved up to higher current values for a nanoelement with larger $M_S$. Summarizing,
the usage of a larger saturation magnetization changes the overall agreement between the
experiment and simulation for the worse.

It is also instructive to compare our results with the simulations performed in the macrospin
approximation in Ref. \onlinecite{Kiselev2003} itself. First of all we note that by performing
such simulations one has to introduce the artificial anisotropy field ${\bf H}_{\rm an}$ (directed 
in this case along the long axis of the ellipsoidal element) in order to reproduce already
the quasistatic hysteresis loop. The magnitude of this field (or the corresponding anisotropy
constant) is usually chosen to obtain the coercivity similar to those measured experimentally.
From the physical point of view this field is intended to describe the average 
magnetocrystalline anisotropy of a polycrystalline element and/or its shape anisotropy. 
The first contribution is negligibly small in samples without a specially induced 
in-plane grain texture (what is probably absent for sputtered Co films).
The second contribution - the shape anisotropy of a thin elliptical nanoelement - can not 
be described adequately by a uniaxial anisotropy term alone, but is in fact a superposition
of a strong 'easy plane' anisotropy due to the demagnetizing field of a thin film and an
additional in-plane anisotropy arising due to a non-circular lateral shape of a nanoelement.
This is one of the reasons why one can expect only a qualitative information from such a
macrospin approximation, even if no domain formation is expected during the magnetization 
reversal.

From this qualitative point of view macrospin simulations presented in 
Ref. \onlinecite{Kiselev2003} demonstrate two interesting features (see Fig. 3 
in Ref. \onlinecite{Kiselev2003}): (i) a frequency jump by the transition from 
a small-angle to a large-angle precession and (ii) an out-of-plane
precession mode with the frequency {\it increasing} with the current strength. 

The first feature is temptingly similar to the frequency jump found for the sample 1 from
Ref. \onlinecite{Kiselev2003} when the current was increased (see Fig. 1d and 1f) and hence was 
proposed by Kiselev et al. as a possible explanation for this experimental observation. 
Our micromagnetic simulations reveal, however, that in a full micromagnetic model this 
jump does {\it not} occur by the transition from a regular to a quasichaotic precession 
regime. The reason for this discrepancy between the macrospin and full micromagnetic model 
is the following. In the macrospin approximation the abrupt transition between the small-angle
(ellipsoidal) and a large angle ('butterfly' or 'shell') precession is accompanied by
the strong decrease of the oscillation frequency because the length of the 'butterfly'
trajectory after the transition is much larger than the length of the elliptical
trajectory before it. Full-scale micromagnetic simulations reveal that the regular
precession regime exists really up to nearly maximal amplitudes of the magnetization
oscillations. This {\it large} amplitudes can not be significantly increased by the 
quasichaotic regime and thus the precession in this regime 'inherits' the frequency
of regular oscillations, and the frequency jump by this transition is absent. Hence
the abovementioned transition can not be the reason for the frequency jump observed
in a real system.

The second feature obtained in the macrospin approximation - the 'out-of-plane' precession
regime - is not observed experimentally \cite{Kiselev2003}. Indeed, our full-scale
simulations show that this precession mode is relatively unstable due  to a slightly
non-collinear magnetization configuration already for the 'minimal' model, where 
neither the polycrystalline structure nor thermal fluctuations are taken into account. 
Either of these two factors disturbs further the homogeneous magnetization structure,
thus nearly wiping out the 'out-of-plane' mode, which intensity is then probably below
the observable threshold.

\subsection{Comparison with experimental data}
\label{subsec:CompExpData}

As it was mentioned in the Introduction, the quality of experimental data presented in 
Ref. \onlinecite{Kiselev2003} is sufficiently high to allow a quantitative theoretical analysis,
but one should of course begin with a qualitative comparison between experimental and
simulated results.

First we note that the presence of several spectral bands shown in Fig. 1f in
Ref. \onlinecite{Kiselev2003} is due to the superposition of signals arising from the
oscillations of the perpendicular and longitudinal magnetization components. This
can be clearly seen from Eq. (\ref{MagResEq2}) and Fig. \ref{fig:ChaoticOscillMinMod}c,
where the magnetoresistance oscillation spectrum obtained using (\ref{MagResEq2})
is presented. Oscillations of the longitudinal ($m_z^{\rm av}$) magnetization component 
produce the signal with the basic (precession) frequency, whereby the magnetization component 
perpendicular to the long axis of the ellipsoid ($m_x^{\rm av}$) is responsible for 
the second harmonic generation: As explained above by the analysis of the 'minimal model' 
spectra (section {\ref{subsec:MinModT0}}), for this type of motion the 
$m_x^{\rm av}$-component oscillates with the frequency roughly twice the precession 
frequency. A weak third harmonics is present in the spectrum of $m_z^{\rm av}$-oscillations 
due to a non-linear precession character for this regime. A quantitative comparison 
of the basic frequency and harmonic intensities is not possible, because the relation 
between corresponding intensities strongly depends on the angle between the external 
field and the ellipsoid axis, which is not known exactly.

The interpretation outlined above is strongly supported by the experimental observation
that the basic frequency band disappeared when the external field was aligned as exact
as possible along the long axis of the elliptical element \cite{KiselevPrivComm}, as
it should be according to Eq. (\ref{MagResEq2}) with $e^h_z = 0$.

The next intriguing feature of the magnetization dynamics measured in Ref. \onlinecite{Kiselev2003}
is the existence of two oscillation regimes - for low (for $2.0 - 2.4$ mA) and high
(from $\approx 2.6$ mA up to $\approx 6.5$ mA) current values. The transition between
these two regimes is accompanied by the drastic increase of the microwave power and by
the frequency jump from $\approx 16$ GHz in a low-amplitude regime down to $\approx 7$ GHz
for large-amplitude oscillations. These features are quite similar to those obtained 
for the macrospin model (see corresponding simulations in Ref. \onlinecite{Kiselev2003}). However,
as explained above, they disappear when this transition is studied using a full 
micromagnetic model with {\it moderate} or {\it absent} random grain anisotropy of 
a polycrystalline elliptical element: First, the frequency during this transition
remains nearly unchanged, and second, the {\it magnitude} of the simulated microwave 
power before the transition is at least as high (and for $T = 0$ even higher) as 
immediately after it (this latter feature is also due to a fact that the amplitude in a 
regular regime just before the transition is by no means small, as it is the case
in the macrospin model).

For this reasons another explanation of the frequency and amplitude relation in these two
regimes is required. Such an explanation can be given assuming that the nanoelement
studied in Ref. \onlinecite{Kiselev2003} had at least a small (but non-negligible) fraction of 
{\it hcp} Co grains (evidences about the coexistence of {\it fcc} and {\it hcp} 
phases in this Co films can be found, e.g., in Ref. \onlinecite{Langer2001}). As shown in 
Fig. \ref{fig:CompDiffRealiz_hcpCo}, polycrystalline 'samples' consisting of 
{\it hcp} Co demonstrate a large variety of possible spectral maps which strongly
depend on the realization of a polycrystalline sample structure. As explained in
section \ref{subsubsec:RandAnisHCP_T0}, this is a natural consequence of a relatively
strong anisotropy of the {\it hcp} grains, which lead to very different magnetization 
dynamics for garins with different directions of the anisotropy axes. This, in turn,
may lead to a spatial 'migration' of the oscillation power by increasing
the current strength. All these circumstances can result in a single 
(Fig. \ref{fig:CompDiffRealiz_hcpCo}b) or multiply (Fig. \ref{fig:CompDiffRealiz_hcpCo}a) 
frequency jumps when increasing current and quite different dependencies of the
spectral maxima frequency on the current strength (compare the behaviour of
broad spectral bands in, e.g., Fig. \ref{fig:CompDiffRealiz_hcpCo}c and d).
The magnitude of the frequency jump can be quite large: we have observed frequency 
changes as large as 10 GHz (in principle even larger jumps are possible:
the maximal oscillation frequency for a {\it hcp} Co grain with $M_S$ of a bulk Co and
the anisotropy axis aligned along the external field is $f_{\rm max} \approx 38$ GHz).
In this connection it is interesting to note that the regular precession frequency
$f_{\rm reg} \approx 17$ GHz obtained for the sample which spectral map is shown in 
Fig. \ref{fig:CompDiffRealiz_hcpCo}b nearly coincides with that measured in the
'small-amplitude' regime in Ref. \onlinecite{Kiselev2003}.

The interpretation proposed above could also explain why the {\it magnitude} of the microwave
power drastically increases after the transition to a quasichaotic behaviour: it might
happen that only a small fraction of a sample (e.g., only {\it hcp} grains with a favourable
orientation of the anisotropy axes) oscillates in a regular ('small-amplitude') regime, whereby in
a quasichaotic regime one observes the oscillations of a major part of a nanoelement
occupied by the {\it fcc} nanocrystals. Our hypothesis is supported further by the
observation that the 'small-amplitude' regime was found not for all experimentally 
studied samples \cite{KiselevPrivComm}.

We would like also to add a few comments concerning the 'out-of-plane' precession 
mode which was {\it not} observed experimentally, despite being predicted by a macrospin
model. As discussed in subsections \ref{subsubsec:RandAnisFCC_T0}, \ref{subsec:SimResT300} 
and \ref{subsec:CompOtherNumSim} above, the absence of this mode in the measured spectra 
can be explained by the joint influence of a polycrystalline sample structure and thermal
fluctuations: both these factors destroy the homogeneous magnetization structure thus
decreasing the amplitude of the 'out-of-plane' mode. In this connection it is worth 
noting that for some {\it hcp}-'samples' we have observed a relatively weak spectral
band corresponding to 'out-of-plane' oscillations (this mode could be identified due to its
very characteristic frequency dependence on the current strength similar to that shown 
in Fig. 3b in Ref. \onlinecite{Kiselev2003}). We have found out that this mode was localized
in a small region of a nanoelement, where the orientations of the grain anisotropy axes
were nearly parallel, thus favouring the collinear magnetization structure during 
the precession. The importance of the polycrystalline structure for the suppression
of the 'out-of-plane' precession could be tested on samples made of materials with 
low crystal grain anisotropy (like Permalloy). For this samples the 'out-of-plane'
mode should be observable, because the nearly absent magnetic anisotropy of Py could
not destroy the homogeneous magnetization configuration required for the existence
of the 'out-of-plane' mode. This conclusion seems to be supported by the recent
experimental observations \cite{KiselevMMM2004}.

As it can be seen from the previous discussion, the extended micromagnetic model which
takes into account a polycrystalline structure of a nanoelement and thermal fluctuations
can explain qualitatively all most important features of the experimental data reported
up to now \cite{Kiselev2003}. However, from the {\it quantitative} point of view, several
discrepancies do exist.

The most important one is the difference between the simulated and measured frequencies
in the quasichaotic regime: at the beginning of this regime the measured frequency
is $f_{\rm mes} \approx 6.5$ GHz (see Fig. 1c in Ref. \onlinecite{Kiselev2003}), whereas
our simulations for elements without a polycrystalline structure and for 
a polycrystalline {\it fcc}-Co give $f_{\rm sim} \approx 9 - 10$ GHz for $a_J$ values where
the transition to a quasichaotic regime is completed (see, e.g., 
Fig. \ref{fig:RegOscillMinMod}c and spectrum for $a_J = 0.38$ in 
Fig. \ref{fig:AllRegimes_T300}). We consider this discrepancy as the most serious one 
because the frequency is the most reliable spectral characteristic obtained both in 
measurements and simulations: whereas other parameters like the line width and intensity
can be artificially altered by many factors, the peak frequency is an inherent signal
feature which is usually reproduced with an accuracy allowing a direct comparison between
simulations and experiment. 

By the discussion of the influence of various factors which could account for the frequency 
difference pointed out above we would like to begin with the reasons having no direct
relation to the spin injection, i.e., with the factors which are not taken into account in
our 'standard' micromagnetic model.

(i) {\it Influence of the underlayer}. In our treatment we have considered a single-layer
elliptical nanoelement, which means that the magnetodipolar interaction with the Co 
underlayer present in a real experiment \cite{Kiselev2003} was not included. 
In principle, such an interaction could indeed decrease the oscillation frequency, 
especially taking into account that the underlayer could be also partially structured 
by the ion milling used in Ref. \onlinecite{Kiselev2003} to produce an elliptical 
nanoelement from the upper Co layer. The elliptical pillar sticking out from 
the underlayer after such a milling would be magnetized in the external
field direction thus producing the stray field in the opposite direction. This latter field
would decrease (on average) the total field acting on the elliptical nanoelement under study
thus decreasing its precession frequency. 

We are not able to simulate this effect quantitatively because the milling depth for 
the underlayer is not known exactly. However, by creating the columnar structure 
studied in Ref. \onlinecite{Kiselev2003} it was aimed to disturb the lower Co layer 
as little as possible, so that the milling depth through the underlayer should be 
maximal several nanometers \cite{KiselevPrivComm}. The stray field from a pillar 
with such a small height acting on the element placed 10 nm above it would be concentrated 
mainly in small regions near those edges of the nanoelement which are perpendicular to 
the external field (or the long ellipsoid axis). Thus such a field could presumably not
lead to a significant decrease of the precession frequency in the quasichaotic regime, where
oscillations of the inner nanoelement regions play a major role. 

To check this conclusion, we have performed simulations of a system consisting of 
an extended underlayer with a thickness 10 nm, an elliptical pillar with the thickness 5 nm
and lateral sizes $130 \times 70$ nm placed on the top of it and a nanoelement with the same 
lateral sizes located 10 nm (thickness of the Cu layer between the two Co layers in Ref.
\onlinecite{Kiselev2003}) above the pillar. Periodic in-plane boundary conditions were 
applied to eliminate the effect of the stray field from a relatively thick extended 
underlayer. We have found (detailed results will be presented elsewhere) that the 
precession frequency both in the regular and in the quasichaotic regime was not 
noticeably influenced by the underlayer with a pillar, as expected according to 
the considerations above.

(ii) {\it Grain texture for an element with a {\it hcp} polycrystalline anisotropy}. The 
already studied influence of the polycrystalline structure for a Co nanoelement with 
{\it hcp} grains could be greatly enhanced if the element would possess a significant
grain texture. In particular, if there would exist an element region with the lateral size
much larger than the element thickness where the grain anisotropy axes are 
approximately parallel to each other and perpendicular to the element plane, then 
the oscillation frequency for a homogeneous precession mode of such a region would be 
$f^{\rm an}_{\rm FMR} = (\gamma/2\pi) \cdot [H_0(H_0 + (4\pi - \beta)M_S)]^{1/2}$ (for
$\beta < 4\pi$), where $\beta = 2K/M_S^2$ is the reduced anisotropy constant 
introduced above. For a {\it hcp} Co with $M_S = 1400$ G and $K = 4.5 \times 10^6$ 
erg/cm$^3$ this means the reduction of the precession frequency by a factor 
$f^{(0)}_{\rm FMR}/f^{\rm an}_{\rm FMR} \approx 1.2$. If we assume that the quasichaotic
precession frequency $f_{\rm ch}$ scales with $M_S$ and $\beta$ 
at least approximately in the same way as the FMR-frequency, the influence of the 
{\it hcp}-anisotropy could decrease $f_{\rm ch}$ from 10 GHz down to $\approx 8.2$ GHz
which would significantly improve the agreement with the experiment. For a Co film
with the reduced $M_S \approx 800$ G the anisotropy $\beta \approx 14$ would be 
even somewhat larger than the shape anisotropy so that the precession with the
frequency close to the free spin value $f_0 = (\gamma/2\pi) \cdot H_0 \approx 5.5$ GHz
and lower (taking into account that $f_{\rm ch} < f^{(0)}_{\rm FMR}$) could be expected.
Unfortunately, as stated above, a highly textured polycrystalline structure should exist
in the sample under study in order to make these arguments applicable. This is unlikely 
for a sputtered Co film (we note in addition, that such an area would probably
lead to an appearance of a significant peak at the corresponding frequency in 
a regular regime). However, this line of the argumentation clearly shows that a detailed 
characterization of a polycrystalline sample structure is necessary when one performs 
measurements on materials with moderate to high magnetocrystalline anisotropy.

(iii) {\it Surface roughness}. In a thin film element with a rough surface a demagnetizing
field due to this roughness would exist, what can also decrease a precession frequency.
Quantitative studies of the roughness effect are possible only when the corresponding 
information about the roughness parameters is available. Here we can only point out 
that a substantial roughness with a characteristic wavelength of the same order as the
element thickness is necessary to create an appreciable demagnetizing field
inside a nanoelement. 

Next we consider in our discussion the possible factors connected to the magnetization
dynamics of a system driven by a spin-polarized current (SPC).

(iv) {\it Effective field induced by the spin-injection}. The existence of such a field
was extensively discussed theoretically (see references in the Introduction), but
only recently this field could be measured in a direct experiment \cite{Zimmler2004} using 
a small asymmetry of the hysteresis loop in a multilayer system in the presence 
of a SPC. For a system very similar to that studied in Ref. \onlinecite{Kiselev2003}
it was shown that the magnitude of a SPC-induced effective field can be calculated as
$H_{\rm SPC} = b_J \cdot j$, where $j$ is the current density and the coefficient $b_J$ was 
found to be $b_J \approx 1.5 \times 10^{-7}$ ${\rm Oe} \times {\rm cm}^2/A$. For 
a typical current value $I = 3$ mA and lateral sizes of the nanoelement under study
$130 \times 70$ nm the current density is $j \approx 4 \cdot 10^7$ A/cm$^2$. 
Hence a SPC-induced field would be $H_{\rm SPC} \approx 6$ Oe which is negligibly 
small.

(v) {\it Non-trivial angular dependence of the Slonczewski torque magnitude $a_J$ and
configuration-dependent damping $\lambda$}. As it was discussed in Sec. \ref{sec:NumSimMeth},
these factors should be considered on equal footing, because the spin-induced torque
and precession damping in multilayer systems subject to a spin-polarized current are
controlled by the processes of the same physical nature. Here we would like only to
mention, that from a qualitative point of view the enhancement of the precession 
damping for large SPC strengths and/or for large angles between the magnetization 
of adjacent layers (see Ref. in Sec. \ref{sec:NumSimMeth}) could indeed decrease the
precession frequency in a quasichaotic regime which is characterized by strong
inhomogeneities of the magnetization structure. The quantitative study of the 
corresponding effect is the subject of a future work.

\section{Conclusion}

In conclusion, we have performed a systematic study of the magnetization dynamics
of a thin single-layer elliptical nanoelement driven by a spin-polarized current flowing 
perpendicular to its plane. Analyzing the dependence of this dynamics on various
magnetic material parameters, we have demonstrated that an extended micromagnetic model 
which takes into account a polycrystalline structure of this element and the influence 
of thermal fluctuations can in principle explain qualitatively almost all features
observed experimentally. At the same time we have shown that an apparent qualitative
agreement between the macrospin model and some experimental findings is due to 
artificial features of the macrospin dynamics, which are absent in the full-scale
micromagnetic model. An important issue concerning the 
quantitative disagreement between the precession frequencies measured experimentally 
and simulated numerically for the quasichaotic regime still requires further clarification.
Analyzing various possible causes of this discrepancy we have shown that the 
most likely candidate to resolve this difficulty is the refined model of the spin-injection
induced dynamics which would include the non-trivial angular dependence of the 
spin-torque term and the dependence of the precession damping on the spin-polarized
current strength and the magnetization configuration of a multilayer system.


\newpage

\newpage

\begin{figure}[tbhp] 
\caption
{
(Color on-line) Spectral power of magnetization oscillations for an elliptical element with a 'reference' 
parameter set for $a_J$ values above the transition from a regular to a quasichaotic regime. 
Oscillation power of $m_x^{\rm av}$ (a) and $m_z^{\rm av}$ (b) magnetization components 
together with the magnetoresistance oscillation power (c) calculated from the relation 
(\ref{MagResEq2}) are presented. A region on the $(a_J - f)$-plane corresponding to
the spectral peaks shown in Fig. \ref{fig:RegOscillMinMod} (regular precession 
regime) is marked as a white rectangle in the panel (b)
} 
\label{fig:ChaoticOscillMinMod}
\end{figure}

\newpage

\begin{figure}[tbhp] 
\centering
{\includegraphics
[scale=0.8, bb=5cm 1cm 15cm 28cm]
{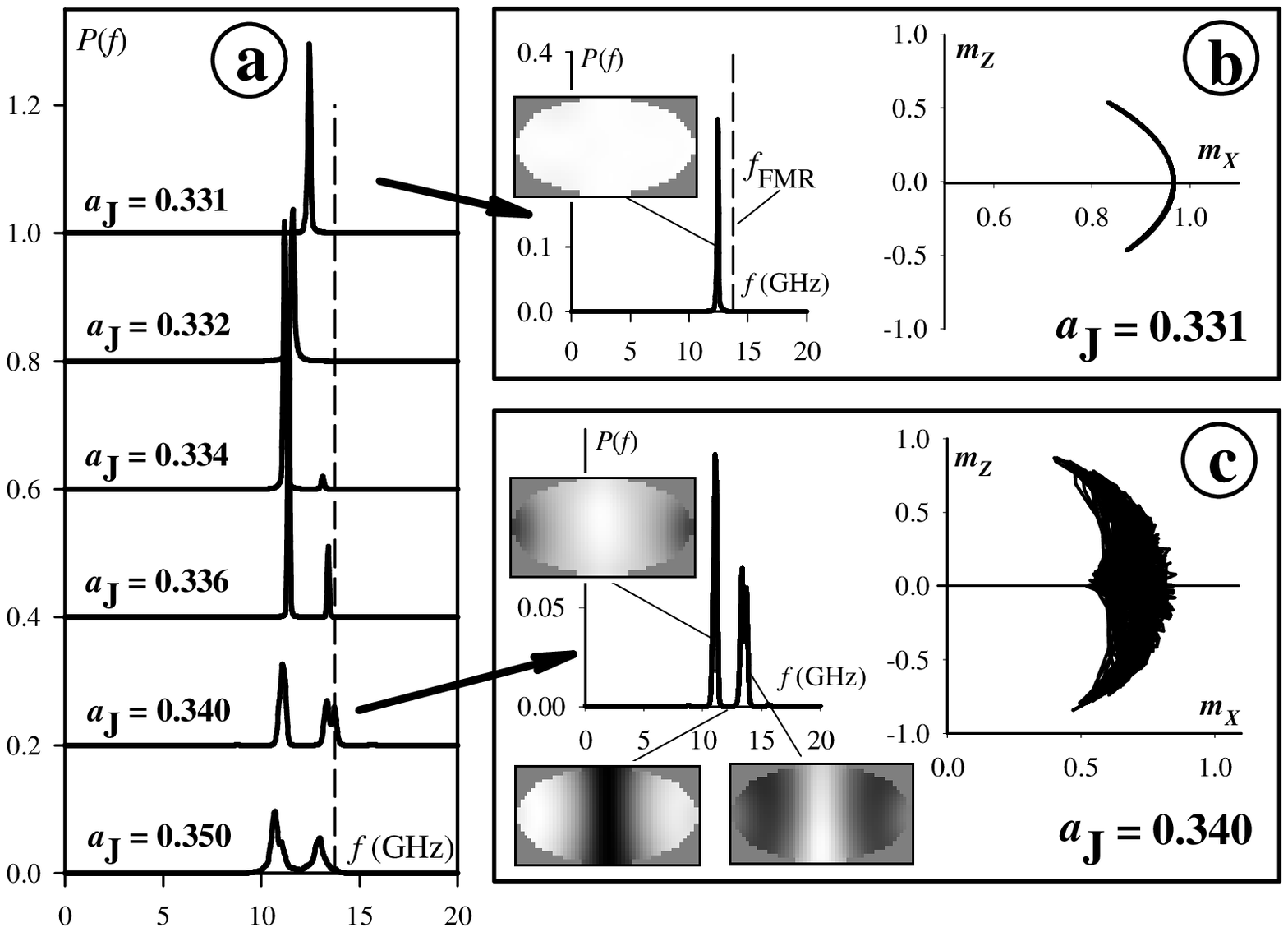}}
\caption
{
Magnetization dynamics for a nanoelement with the exchange constant reduced down to 
$A = 2.0 \cdot 10^{-6}$ erg/cm (other parameters are as for a 'reference' parameter set)
represented in the same way as in Fig. \ref{fig:RegOscillMinMod}. Note that the maximal
oscillation power in a regular precession regime is much smaller than in 
Fig. \ref{fig:RegOscillMinMod}.
}
\label{fig:RegOscillA2} 
\end{figure}

\newpage

\begin{figure}[tbhp] 
\centering
{\includegraphics
[scale=0.8, bb=5cm 1cm 15cm 28cm]
{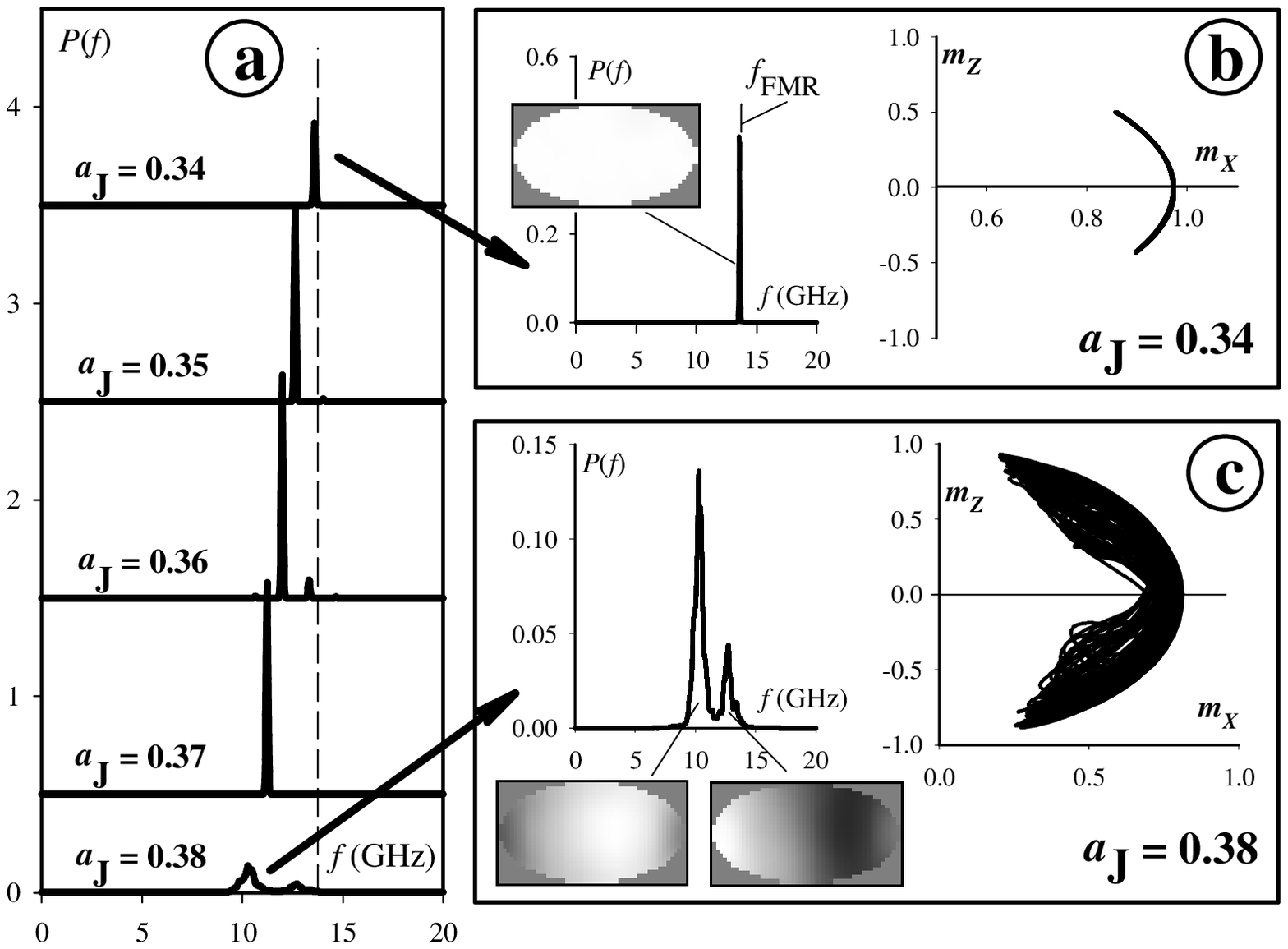}}
\caption
{
The same as in Fig. \ref{fig:RegOscillMinMod} for a nanoelement with the 'reference'
parameter set and the random magnetic anisotropy corresponding to a polycrystalline
{\it fcc} Co film structure with the average grain size $\langle D \rangle = 10$ nm.
}
\label{fig:RegOscill_fccAnis} 
\end{figure}

\newpage

\begin{figure}[tbhp] 
\caption
{
(Color on-line) Comparison of the magnetoresistance power spectra in a quasichaotic regime for
nanoelements with the 'reference' parameter set (a), the 'reference' parameter set 
except the reduced exchange stiffness constant $A = 2.0 \cdot 10^{-6}$ erg/cm (b), 
the 'reference' parameter set and a polycrystalline structure with average grain 
size $\langle D \rangle = 10$ nm) (c). For the last case the random orientations
of anisotropy axes for different grains and the {\it fcc} Co structure (cubic anisotropy
of grains with $K_1^{\rm cub} = 6.0 \cdot 10^5$ erg/cm$^3$) were assumed.  
It can be seen that both the reduction of the exchange stiffness constant and
inclusion of a random anisotropy lead to (i) the expansion of the $a_J$ region 
where quasichaotic oscillations exist and (ii) decrease of the out-of-plane 
oscillation power.
}
\label{fig:CompChaoticOscill_A3_A2_fccCo} 
\end{figure}

\newpage

\begin{figure}[tbhp] 
\caption
{
(Color on-line) Maps of the magnetoresistance oscillation power for a nanoelement with 
the 'reference' parameter set and a polycrystalline structure with {\it hcp}{\it hcp} Co grains.
The grains with $\langle D \rangle = 10$ nm have a uniaxial anisotropy 
($K_1^{\rm un} = 4.5 \cdot 10^6$ erg/cm$^3$) and randomly oriented anisotropy 
axes. Maps (a)-(d) represent oscillation power spectra for nanoelements with
the same macroscopic parameters, but {\it different} realizations of a random
polycrystalline structure.
}
\label{fig:CompDiffRealiz_hcpCo} 
\end{figure}

\newpage

\begin{figure}[tbhp] 
\centering
{\includegraphics
[scale=0.8, bb=5cm 1cm 15cm 28cm]
{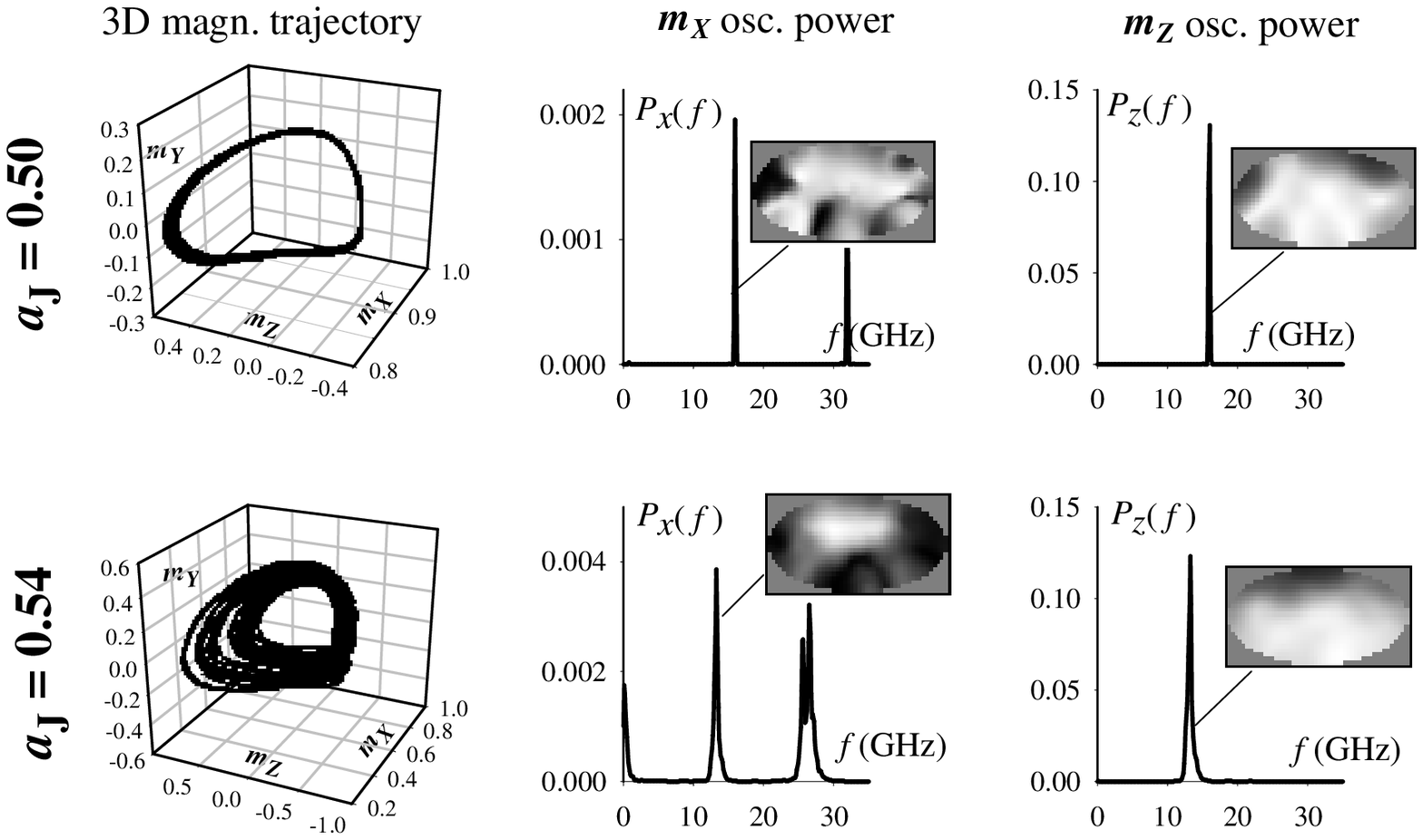}}
\caption
{
Magnetization trajectories (a,d) and oscillation power spectra for $m_x^{\rm av}$ (b,e) 
and $m_z^{\rm av}$ (c,f) magnetization components for a random polycrystalline structure 
realization marked as (b) in Fig. \ref{fig:CompDiffRealiz_hcpCo}. The graphs in the upper 
row are for the current value $a_J = 0.50$ before the transition from a regular to 
a quasichaotic behaviour, the lower row shows results for $a_J = 0.54$ which is immediately 
after this transition. Maps in the graphs (b),(c),(e) and (f) display the spatial
distribution of the oscillation power for the corresponding magnetization components.
}
\label{fig:3Dtraj_SpecMxMz_hcpCo} 
\end{figure}

\newpage

\begin{figure}[tbhp] 
\caption
{
(Color on-line) Magnetization dynamics of a polycrystalline ({\it fcc} Co) nanoelement 
at $T = 300$ K: transition from a regular to a quasichaotic behaviour (a) and the map 
of the magnetoresistance oscillation power for the whole $a_J$-range (b). Comparison 
of the latter map to the analogous map for $T = 0$ in 
Fig. \ref{fig:CompChaoticOscill_A3_A2_fccCo}(c) shows that thermal fluctuations lead 
to (i) an appearance of an extended region of a regular precession and (ii) a strong 
power decrease of the 'out-of-plane' oscillations.
}
\label{fig:AllRegimes_T300}
\end{figure}

\end{document}